\documentclass{aastex631}

\usepackage{epstopdf}
\usepackage{natbib}
\usepackage{graphicx}
\usepackage{CJK}

\usepackage[percent]{overpic}
\usepackage{times}
\usepackage{graphicx}
\usepackage{amsmath}

\usepackage{subfigure}
\usepackage{natbib}
\usepackage{txfonts}
\usepackage{journals}
\usepackage{xcolor}
\usepackage[utf8]{inputenc}
\usepackage[overload]{empheq}
\usepackage{verbatim}

\submitjournal{ApJS}
\begin{document}
\begin{CJK*}{UTF8}{gbsn}

\title{A High-Resolution Stellar Extinction Map of M31 from PHAT and PHAST Photometry} 

\author[0000-0002-2473-9948]{Mingxu Sun (孙明旭)}
\affiliation{Department of Physics,
               Hebei Key Laboratory of Photophysics Research and Application,
               Hebei Normal University,
               Shijiazhuang 050024, P.\,R.\,China}

\author[0000-0003-2472-4903]{Bingqiu Chen (陈丙秋)}
\affiliation{South-Western Institute for Astronomy Research,
               Yunnan University,
               Kunming 650500, P.\,R.\,China}

 \author[0000-0003-3860-5286]{Yuxi Wang (王钰溪)}
 \affiliation{Department of Astronomy,
                College of Physics and Electronic Engineering,
                Qilu Normal University,
                Jinan 250200, P.\,R.\,China}

 \author[0000-0003-4195-0195]{Jian Gao (高健)}
 \affiliation{Department of Astronomy,
                Beijing Normal University,
                Beijing 100875, P.\,R.\,China}

\author[0009-0001-4609-7069]{Yuhui Chen (陈玉辉)}
\affiliation{Department of Physics,
               Hebei Key Laboratory of Photophysics Research and Application, 
               Hebei Normal University,
               Shijiazhuang 050024, P.\,R.\,China}
               
 \author[0000-0003-1359-9908]{Wenyuan Cui (崔文元)}
 \affiliation{Department of Physics,
                Hebei Key Laboratory of Photophysics Research and Application, 
                Hebei Normal University,
                Shijiazhuang 050024, P.\,R.\,China}        
                
\correspondingauthor{Bingqiu Chen; Mingxu Sun; Yuxi Wang}
\email{bchen@ynu.edu.cn; mxsun@hebtu.edu.cn; yuxiwang@qlnu.edu.cn}

\date{Accepted 2026 November 19. Received 2026 October 10; in original form 2026 June 2}

\label{firstpage}

\begin{abstract}
High-resolution extinction maps of resolved external galaxies, such as M31, are crucial not only for accurately correcting dust extinction and reddening toward individual sources within the galaxy, but also for characterizing the distribution and evolution of interstellar dust in external galactic environments. We aim to construct a high-resolution and large-coverage stellar-traced extinction map of M31, based on multi-band photometric data from the HST PHAT and PHAST surveys. Using a stellar template library built from the PARSEC stellar evolution models and adopting the average extinction law of M31, we apply a matching method analogous to the SPar algorithm to derive $A_V$ values for $\sim 65$ million stars. The extinction map is then generated using a HEALPix adaptive-resolution scheme. The resulting map covers approximately two-thirds of the star-forming disk of M31, with a maximum spatial resolution of $1.61''$. The median extinction across the map is $A_V = 1.24\,\mathrm{mag}$. Comparison with previously published stellar- and dust-emission-based extinction maps shows broadly similar large-scale morphology, while differences remain in the extinction values. The extinction map further reveals fine-scale features such as inter-arm filaments, supernova-driven cavities, and spur-like structures. The map has been publicly released through the NADC platform and is expected to serve as an important observational basis for extinction correction in subsequent astronomical observations and for related M31 scientific studies.

\end{abstract}


\section{Introduction} \label{sec:intro}

Correcting for interstellar dust extinction is essential for accurately determining stellar properties, star formation histories, and distance scales in extragalactic systems \citep{Calzetti2001, Galliano2018, Salim2020}. Dust grains absorb and scatter starlight, significantly modifying the observed appearance of disk galaxies, particularly at optical and ultraviolet wavelengths, while approximately 30\% of the stellar energy is reradiated by dust in the infrared \citep{Disney1989, Xilouris1999, Calzetti2001}. In recent decades, advances in mid- and far-infrared observational facilities have established dust as an independent research topic \citep{Savage1979, Draine2003}, and extinction mapping has proven to be one of the most effective tracers of cold, dense molecular gas \citep{Dickman1978, Lada1994, Lombardi2001}. At the same time, high-resolution extinction maps themselves provide key observational constraints on dust grain physics and theories of interstellar medium (ISM) evolution \citep{Kainulainen2009}. M31, the Andromeda galaxy, lies at a distance of approximately $780\,\mathrm{kpc}$ from the Milky Way \citep{McConnachie2005} and is the nearest large spiral galaxy, as well as the only galaxy in the Local Group similar to the Milky Way in mass, morphology, and metallicity \citep{Driver2007}. The construction of a high-resolution extinction map of M31 is therefore essential for advancing core scientific objectives related to stellar physics, galactic structure, and chemical evolution within this system.

Existing extinction maps of M31 have been derived mainly through two approaches. With respect to methods based on dust thermal emission, which have been widely used to map dust distributions in nearby galaxies \citep[e.g.,][]{Barmby2006, Groves2012, Fritz2012}, \citet{Draine2014} employed Spitzer and Herschel data to construct a dust mass surface density map of M31, which was then converted to a $V$-band extinction map using a dust model. However, follow-up studies \citep{Dalcanton2015,Wang2025} have reported that the extinction values derived by \citet{Draine2014} are generally higher than their respective stellar-based extinction estimates. Differences between the two approaches may reflect both assumptions about dust grain emissivity and selection effects in stellar maps, which can preferentially sample less extinguished sightlines when subpixel dust structure obscures the most heavily reddened stars.

With respect to methods based on stellar photometric tracers, existing studies have differed in their emphasis on spatial coverage and resolution. For localized high-resolution measurements, which have greatly benefited from HST's unparalleled angular resolution \citep[e.g.,][]{Brown2003, Brown2006, Rich1996}, \citet{Dong2016} used Hubble Space Telescope (HST) Wide Field Camera 3 (WFC3) and Advanced Camera for Surveys (ACS) observations in ten bands (from the near-ultraviolet $2700\,\text{\AA}$ to the near-infrared $1.5\,\mu\mathrm{m}$) to construct an extinction map of the central $\sim180''$ ($\sim680\,\mathrm{pc}$) of the M31 bulge through pixel-by-pixel spectral energy distribution (SED) fitting. The map reaches an angular resolution of $\sim0.5''$ (about $2\,\mathrm{pc}$) and an extinction uncertainty of only $\delta A_V\sim0.05\,\mathrm{mag}$, but covers only a very small region at the center of the M31 bulge. For disk-scale measurements, \citet{Dalcanton2015} used HST near-infrared photometry from the PHAT survey to construct a $25\,\mathrm{pc}$-resolution extinction map of M31 by modeling the color--magnitude diagram (CMD) of red giant branch (RGB) stars, covering about one-third of the star-forming disk; however, this map has not been publicly released, which limits its wider use. For large-scale measurements, \citet{Wang2025} employed ground-based multi-band data from the United Kingdom Infrared Telescope (UKIRT) Wide Field Camera (WFCAM; \citealt{Irwin2013}), Panoramic Survey Telescope and Rapid Response System 1 (Pan-STARRS1; \citealt{Chambers2016}), and Gaia Data Release 3 (DR3; \citealt{Gaia2023}), using M31 member stars as tracers, to construct a large-scale extinction map covering about $10\,\mathrm{deg}^2$, reporting a mean extinction of $A_V=1.17\,\mathrm{mag}$ and isolating the Galactic foreground extinction component ($A_V\approx0.185\,\mathrm{mag}$). However, due to the limited number of tracers available from ground-based observations, the spatial resolution of this map is about $50''$, which cannot resolve inter-arm filamentary networks, supernova-driven cavities, and spur-like structures along the spiral-arm edges. These features are key probes for studying physical processes such as ISM turbulence, energetic feedback, and dust evolution.

We aim to construct a stellar-traced extinction map that combines high spatial resolution ($\lesssim 2''$) with large-area coverage of most of the star-forming disk of M31, using the unprecedented photometric data from the Hubble Space Telescope (HST). The PHAT survey covers approximately one-third of the star-forming disk of M31 in six bands from F275W to F160W, with the final catalog containing $1.38 \times 10^8$ stars \citep{Dalcanton2012,Williams2014,Williams2023}. The Panchromatic Hubble Andromeda Southern Treasury (PHAST; \citealt{Chen2025}) survey, which extends PHAT to the southern disk, provides photometric data for over $9 \times 10^7$ stars in four bands (F275W, F336W, F475W, and F814W). Together, the two surveys cover approximately two-thirds of the star-forming disk of M31, providing the data foundation for high-resolution extinction measurements over most of the disk. In addition, \citet{Wang2022} combined multi-band observations to derive the average extinction curve of M31 ($R_V \approx 3.51$), providing an extinction law tailored to M31 that avoids the systematic errors that could arise from adopting the Galactic extinction law to convert HST multi-band photometry into $A_V$. Furthermore, the Stellar parameter from multiband photometry (SPar; \citealt{Sun2023,Sun2025}) algorithm based on stellar template matching has been systematically validated in Galactic extinction measurements, providing a mature framework for deriving single-star $A_V$ values from large-scale multi-band photometry.

In this work, we use the combined multi-band photometry from PHAT and PHAST and apply a stellar matching method similar to the SPar algorithm. Combined with a stellar template library built from the PARSEC stellar evolution models (PAdova and TRieste Stellar Evolution Code; \citealt{Bressan2012}) and the average extinction law of M31 from \citet{Wang2022}, we derive $A_V$ values for 64.8 million high-quality stars. On this basis, we use a HEALPix adaptive-resolution scheme (base resolution $6.44''$, dynamically refined to $3.22''$ and $1.61''$ according to local source density) to construct a multi-resolution extinction map with a maximum spatial resolution of $1.61''$ (about $6\,\mathrm{pc}$ at the distance of M31), covering about two-thirds of the star-forming disk. This paper is organized as follows. Section~\ref{sec:dat} introduces the PHAT and PHAST survey data and the sample selection criteria. Section~\ref{sec:emp} describes the stellar template matching method and the HEALPix-based extinction map generation scheme. Section~\ref{sec:dis} presents the results and provides comparisons with previous studies. Section~\ref{sec:sum} summarizes the work.

\section{Data}\label{sec:dat}

The HST photometric data used in this work come from two complementary survey programs: the PHAT survey, which covers the northern disk of M31, and the PHAST survey, which covers the southern extension. Together, the two surveys provide photometric data for more than 228 million stars.

The PHAT survey is an HST multi-cycle treasury program that uses the WFC3/UVIS, WFC3/IR, and ACS/WFC instruments to perform mosaic imaging of approximately $0.5\,\mathrm{deg}^2$ of the northern star-forming disk of M31 in six bands: near-ultraviolet F275W and F336W, optical F475W and F814W, and near-infrared F110W and F160W \citep{Dalcanton2012}. The survey was conducted between 2010 and 2013. \citet{Williams2014} released the first multi-band photometric catalog, containing ultraviolet to near-infrared photometry for 117 million stars. \citet{Williams2023} reprocessed all imaging data using an improved DOLPHOT point-spread function (PSF) photometry pipeline \citep{Dolphin2000, Dolphin2005, Dolphin2016} and released the final legacy catalog, which contains 138 million stars, an increase of 18\% over the previous version, with up to a 40\% increase in star counts in low-density regions. This catalog provides a nearly seamless map of stellar populations: old stellar populations ($\gtrsim 1$--$2\,\mathrm{Gyr}$) are smoothly distributed, while young stellar populations show a highly structured distribution.

The PHAST survey is the southern extension of PHAT and provides systematic imaging of approximately $0.45\,\mathrm{deg}^2$ in the southern star-forming disk of M31 \citep{Chen2025}. Observations were carried out between 2021 and 2024 using ACS/WFC and WFC3/UVIS in four optical and near-ultraviolet bands (F475W, F814W, F275W, and F336W). Compared with PHAT, PHAST does not include near-infrared bands. By stacking all exposures, the survey reaches a 50\% completeness limit of F475W $\sim 27.7\,\mathrm{mag}$ in the low-density regions of the outer disk, and F475W $\sim 26.0\,\mathrm{mag}$ in the crowded high-surface-brightness regions near the bulge. The final catalog provides DOLPHOT PSF photometric data for more than 90 million stars. The combined PHAT and PHAST data cover approximately two-thirds of the star-forming disk of M31.

We adopt the photometric catalogs of \citet{Williams2023} and \citet{Chen2025} and apply the following selection criteria. First, we remove measurements with photometric errors larger than $0.1\,\mathrm{mag}$ in the F110W, F160W, F475W, or F814W bands, larger than $0.2\,\mathrm{mag}$ in the F275W or F336W bands, or with a signal-to-noise ratio below 5 in any band. Sources must then have valid measurements in at least two of the six PHAT bands (F110W, F160W, F275W, F336W, F475W, and F814W) or in at least two of the four PHAST bands (F275W, F336W, F475W, and F814W). After these selections, the retained PHAT catalog contains $68{,}699{,}907$ sources and the retained PHAST catalog contains $29{,}690{,}596$ sources.

\section{Methods}\label{sec:emp}

Based on the data described in Section~\ref{sec:dat}, the extinction map is constructed in two stages: $A_V$ values for individual stars are first derived through stellar template matching, and these values are then combined into a two-dimensional extinction map using the HEALPix scheme.

\subsection{Derivation of Single-Star Extinction Values}

We adopt a stellar template-matching approach to derive $A_V$ for individual stars. This method follows the core logic of the SPar algorithm developed by \citet{Sun2023,Sun2025}, using $\chi^2$ minimization as the matching criterion, where the best-fit extinction is found by comparing the observed multi-band photometry with the magnitudes predicted by stellar templates. SPar has been systematically validated in stellar-parameter measurements based on the SkyMapper Southern Survey (SMSS; \citealt{Wolf2018,Onken2019}) of the Milky Way, and its basic matching strategy---iterating over templates and trial $A_V$ values, computing $\chi^2$, and selecting the minimum---is retained in this work. Relative to the original implementation of SPar, two adjustments are made in this work. First, SPar uses an empirical stellar library constructed from the LAMOST survey, whereas we use PARSEC stellar evolution models to construct a uniform template grid that covers the stellar-parameter and luminosity ranges required for this study. Second, SPar adopts a Galactic extinction law, whereas this work uses the average extinction law of M31 derived by \citet{Wang2022} from multi-band observations ($R_V\approx3.51$). Spatial variations in the M31 extinction curve may introduce an additional systematic uncertainty in $A_V$. Tests using representative curves corresponding to the 16th--84th percentile range of the \citet{Wang2022} dust-size parameter give 68th-percentile values of $|\Delta A_V|$ of about $0.2\,\mathrm{mag}$ in the sampled pixels. Because local extinction curves are available only along discrete sightlines, we retain the mean M31 curve as a common reference throughout the map.

The stellar template library used in this work is constructed from the PARSEC stellar evolution models \citep{Bressan2012}. Stars are selected at five age points ($50\,\mathrm{Myr}$, $500\,\mathrm{Myr}$, $1\,\mathrm{Gyr}$, $5\,\mathrm{Gyr}$, and $10\,\mathrm{Gyr}$) spanning stellar populations from young to old. The PARSEC models classify stellar evolution into several stages, including the main sequence (MS, label 1), subgiant branch (SGB, label 2), RGB (label 3), core helium burning (CHEB, label 4), CHEB1 (label 5), CHEB2 (label 6), early asymptotic giant branch (EAGB, label 7), and thermally pulsing asymptotic giant branch (TPAGB, label 8). Because stars in the RGB, CHEB, and EAGB phases are brighter and remain detectable by HST at the distance of M31 ($\sim 780\,\mathrm{kpc}$), only stars with labels 3 (RGB), 4 (CHEB), and 7 (EAGB) are used to build the template library. The final library contains $3773$ stellar templates, each providing the absolute magnitude $M_x$ in every band, where the subscript $x$ denotes any given band. The effective temperature and metallicity associated with the best-fitting PARSEC template are used only as coordinates of the template grid in the present analysis and are not treated as independent stellar-parameter measurements.

To determine the conversion between extinction values $A_x$ in different bands and $A_V$, we convolve ATLAS9 stellar atmosphere models \citep{Kurucz1993,Castelli1997,Castelli2003} with the throughput curves of the HST filters and combine the result with the average extinction law of M31 from \citet{Wang2022} to compute the ratio $A_x / A_V$ at different effective temperatures $T_{\rm eff}$ and extinctions $A_V$. Two-dimensional linear interpolation then yields $A_x / A_V$ as a continuous function of $T_{\rm eff}$ and $A_V$. In the subsequent matching procedure, for any given template $T_{\rm eff}$ and trial $A_V$, the extinction values $A_x$ in each band can be obtained directly from this function.

In the template-matching stage, for each star we iterate over all templates in the library and a series of trial $A_V$ values: from $-0.20$ to $10.0\,\mathrm{mag}$ in steps of $0.02\,\mathrm{mag}$, and from $10.05$ to $15.0\,\mathrm{mag}$ in steps of $0.05\,\mathrm{mag}$. For a given template $i$ and trial $A_V$, the model apparent magnitude in band $x$ is written as $m_x^{\rm mod}=M_x(i)+A_x(A_V)+\mu_{\rm M31}$. All stars are assigned the same fixed M31 distance modulus, $\mu_{\rm M31}\simeq24.46\,\mathrm{mag}$ (corresponding to $\sim780\,\mathrm{kpc}$), which is kept fixed and is not treated as a free parameter in the fitting.

For each star, the template and trial $A_V$ that minimize the error-weighted residual sum of squares,
\[
\chi_{\rm raw}^{2}
=
\sum_{x=1}^{N}
\left[
\frac{m_x^{\rm obs}-m_x^{\rm mod}}{\sigma_x}
\right]^2,
\]
are adopted as the best fit, and the corresponding $A_V$ is taken as the extinction value of the star. Here, $m_x^{\rm obs}$ and $m_x^{\rm mod}$ are, respectively, the observed and modeled magnitudes in band $x$, $\sigma_x$ is the observational uncertainty, and $N$ is the number of valid bands used in the fitting. For PHAT stars, $N$ is at most 6 (F110W, F160W, F275W, F336W, F475W, and F814W); for PHAST stars, $N$ is at most 4 (F275W, F336W, F475W, and F814W). For fitting-quality selection, we define $Q=\chi_{\rm raw}^{2}/(N-1)$ and adopt $Q\leq1$ as an empirical quality-selection criterion; here, $N-1$ is not interpreted as the formal number of effective degrees of freedom. This criterion measures the agreement between the observed and modeled photometry but does not guarantee that stellar temperature and extinction are uniquely constrained, particularly for two-band fits.

To quantify the individual-star $A_V$ uncertainties in the quality-selected sample, we evaluate Gaussian likelihoods over the full stellar-template and extinction grid using the catalogue photometric errors. We sum the likelihoods over the equally weighted discrete stellar templates and normalize the resulting $A_V$ probability distribution, adopting its standard deviation as the extinction uncertainty. The median uncertainty is $0.431\,\mathrm{mag}$ for the full sample and $0.477$, $0.420$, and $0.317\,\mathrm{mag}$ for stars with two, three, and at least four valid bands, respectively. These uncertainties are estimated using the adopted stellar templates, extinction law, and catalogue photometric errors and do not fully account for model biases, extinction-law variations, or systematic photometric offsets.

\subsection{Construction of the Extinction Map}

\begin{figure*}[htbp]
\centering
\includegraphics[width=\textwidth]{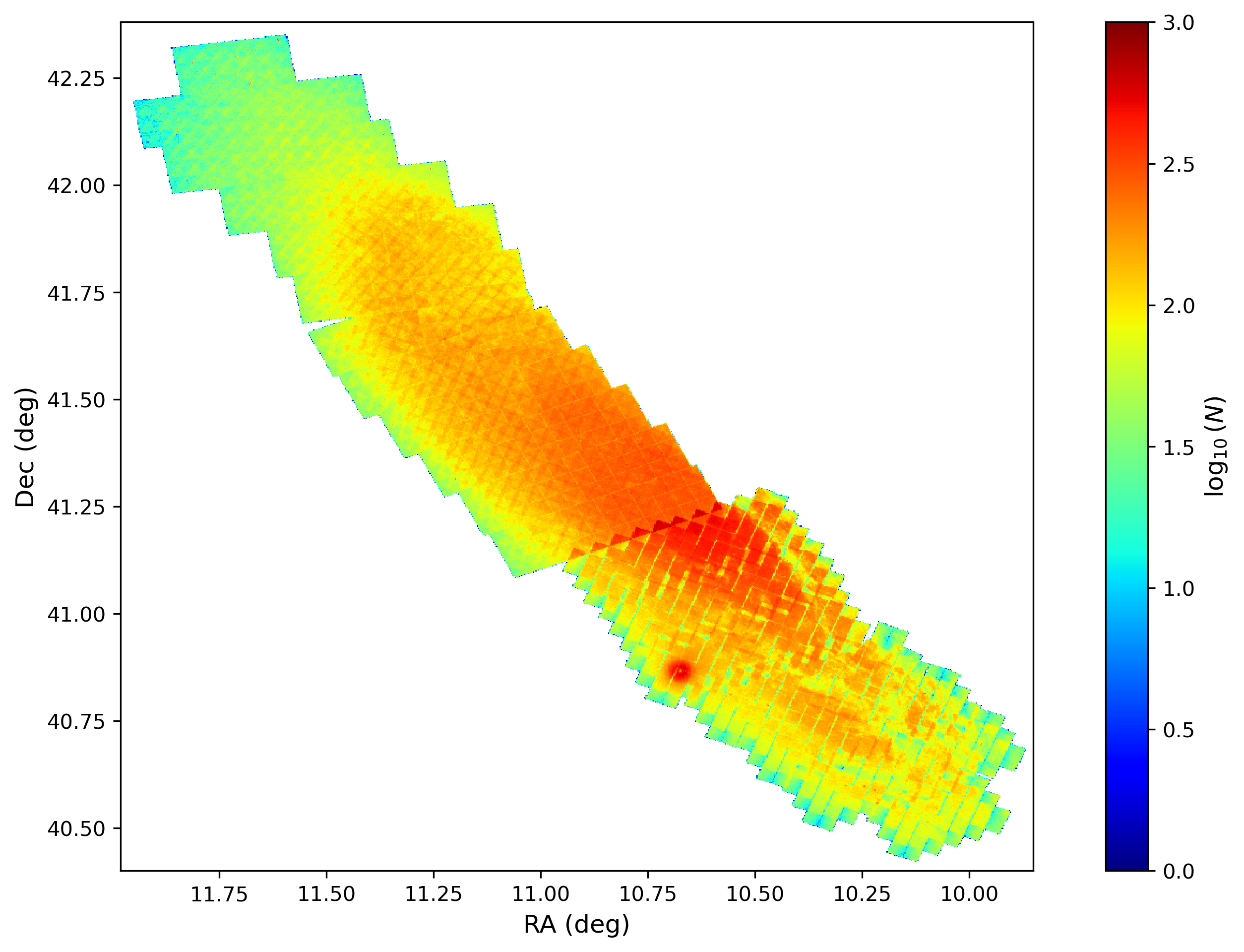}
\caption{Spatial distribution of the number of remaining tracers (after applying the $Q\leq1$ selection and $A_V>0.1\,\mathrm{mag}$ cut) per pixel at the HEALPix base resolution ($nside=2^{15}$, pixel size $6.44''$). The color encodes the source counts on a logarithmic scale.}

\label{mn}
\end{figure*}

After deriving $A_V$ for individual stars, we use the HEALPix scheme \citep{Gorski2005,Zonca2019} to construct the two-dimensional extinction map of M31. The base resolution is set to $nside = 2^{15}$, corresponding to a pixel angular size of $6.44''$ (approximately $24.5\,\mathrm{pc}$ at the distance of $780\,\mathrm{kpc}$). Figure~\ref{mn} shows the spatial distribution of the number of selected background tracers per pixel at this resolution, with a median of $104$ and a maximum of $736$. The source density increases markedly toward the center of M31.

To convert the individual-star $A_V$ measurements into line-of-sight extinction, we adopt the thin-dust-layer assumption used by \citet{Dalcanton2015} and \citet{Wang2025}: the scale height of the M31 dust disk ($\sim0.24\,\mathrm{kpc}$) is much smaller than the characteristic scale height of the stellar disk ($\sim1.1$--$2.8\,\mathrm{kpc}$). Along a given line of sight, stars can therefore be separated into those lying in front of the dust layer, which are largely unaffected by internal M31 extinction, and those lying behind it and subject to extinction. The extinction map derived in this work represents the total line-of-sight extinction toward M31 stars, including contributions from both the Milky Way foreground and M31 itself. For each pixel, we first calculate the median fitted $A_V$ of the stars that satisfy the $Q$ quality criterion, and then select as background-star candidates those with $A_V>0.1\,\mathrm{mag}$ and $A_V$ above this median. When at least four background-star candidates are available, their median $A_V$ is adopted as the extinction value of the pixel. We also tested the sensitivity of the extinction map to the adopted $A_V$ threshold. At locations that are valid in both maps and have the same effective resolution, more than 96\% of the comparable positions retain the same extinction values when the threshold is changed from the adopted $0.10\,\mathrm{mag}$ to 0, 0.05, or $0.15\,\mathrm{mag}$. However, pixel values, effective resolution, or spatial coverage may still change in some low-extinction regions.

To achieve higher spatial resolution in regions of high source density while preserving statistical reliability in low-density regions, an adaptive-resolution scheme is adopted. For pixels containing more than 40 selected background stars, the nested structure of HEALPix is used to dynamically subdivide the pixels into four child pixels, increasing $nside$ successively to $2^{16}$ ($3.22''$, approximately $12.2\,\mathrm{pc}$) and $2^{17}$ ($1.61''$, approximately $6.1\,\mathrm{pc}$). At each resolution level, only pixels containing at least four selected background stars are retained. After filtering and adaptive refinement, the final extinction map contains $39{,}147$, $640{,}370$, and $1{,}023{,}028$ native pixels at $nside=2^{15}$, $2^{16}$, and $2^{17}$, respectively.

\section{Results and Discussion}\label{sec:dis}

\begin{figure*}[htbp]
\centering
\includegraphics[width=\textwidth]{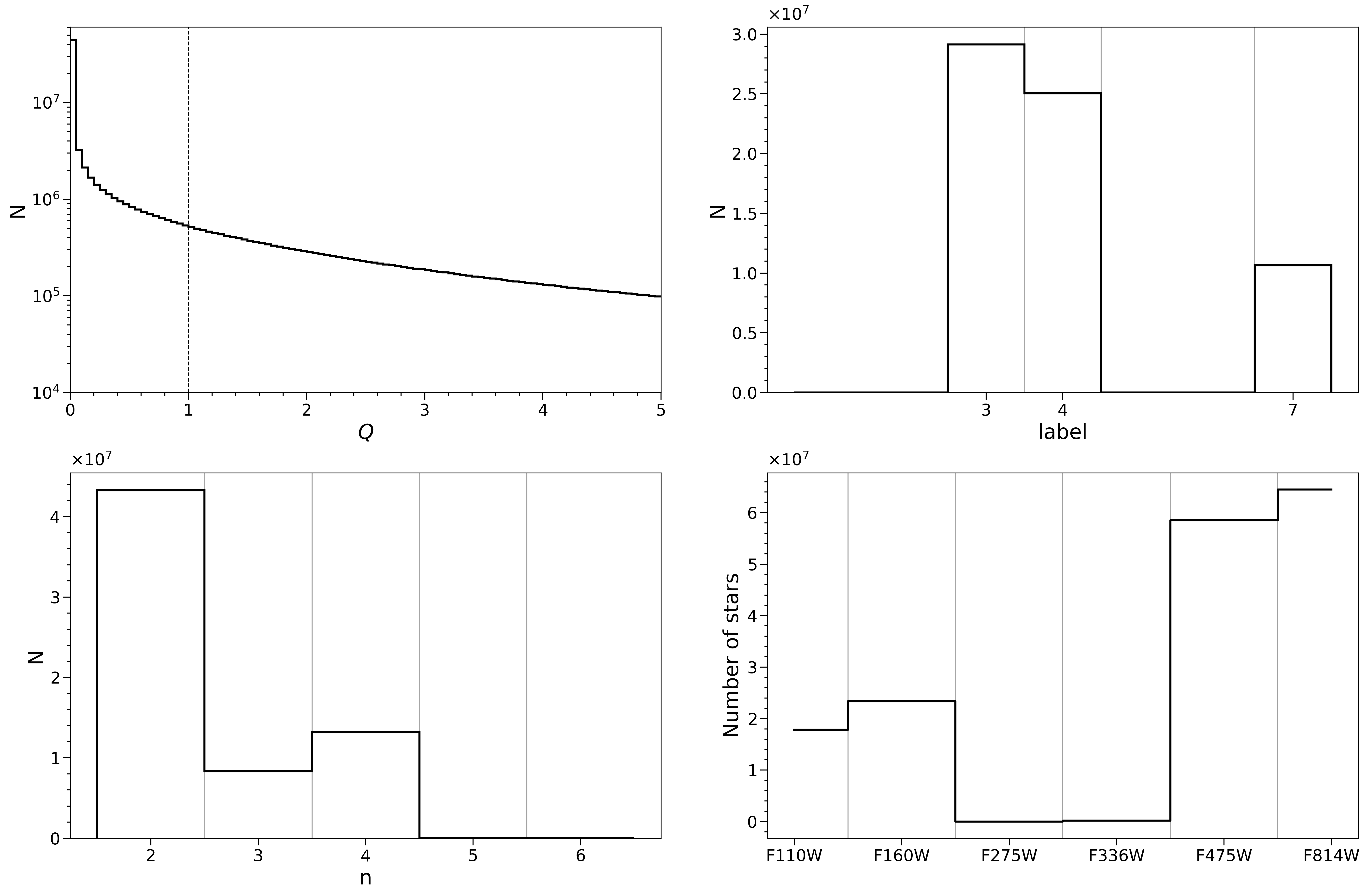}
\caption{The upper-left panel shows the distribution of $Q=\chi^2_{\rm raw}/(N-1)$ for the full fitting sample of 98,390,503 stars. The dashed line marks the selection threshold at $Q=1$, leaving 64,840,973 stars. The upper-right panel shows the distribution of the PARSEC evolutionary-stage labels of the best-fit templates for the $Q\leq1$ subsample (label 3: RGB; label 4: CHEB; label 7: EAGB). The lower-left panel shows the distribution of the actual number of bands $n$ used in the photometric fitting of the $Q\leq1$ stars. The lower-right panel shows the measurement coverage in each band, i.e., the number of stars with valid photometry in that band.}

\label{distr}
\end{figure*}

Using the method described above, we derive $A_V$ values for $98{,}390{,}503$ stars from the PHAT and PHAST survey data, the largest sample yet for stellar extinction measurements in M31. Figure~\ref{distr} presents the $Q$ distribution of the full sample before the fitting-quality selection, together with the distributions of the best-fitting stellar types and valid photometric bands in the selected sample. The overall $Q$ distribution is strongly concentrated at low values. We adopt $Q\leq1$ as an empirical fitting-quality criterion, retaining $64{,}840{,}973$ stars (approximately 65.9\%) for constructing the extinction map. We adopt this quality-selection criterion to limit large discrepancies between the observed photometry and the adopted stellar templates. The template library is constructed from PARSEC theoretical models, whose predicted magnitudes and effective-temperature scales may contain systematic uncertainties. In addition, HST photometry of M31 stars is affected by observational challenges such as crowding, photometric uncertainties, and limited completeness, particularly in dense stellar fields. These factors may influence the template matching and the resulting extinction estimates. The $Q\leq1$ criterion helps exclude fits with relatively large normalized residuals. However, $Q$ is used to assess the agreement between the observed and modeled photometry and does not guarantee that the fitted extinction is uniquely determined or free from systematic biases. In particular, temperature and extinction remain degenerate for stars with only two valid photometric bands.

Within the high-quality subsample, $13{,}222{,}395$ stars (about $20\%$) have valid measurements in four or more independent bands. Multi-band coverage is essential for separating extinction from intrinsic stellar parameters: when only two bands (i.e., a single color) are available, extinction and temperature become fully degenerate in photometric space and cannot be uniquely determined. UV--optical--near-infrared coverage provides additional color and magnitude constraints that can reduce this degeneracy, although the inferred extinction remains dependent on the adopted templates. The lower-left and lower-right panels of Figure~\ref{distr} show the band coverage of the high-quality subsample: most stars have independent measurements in $2$--$4$ bands, and F475W and F814W show the highest coverage because they are included in both surveys. According to the PARSEC evolutionary-stage labels of the best-fit templates (upper-right panel of Figure~\ref{distr}), $29{,}140{,}857$ RGB stars (label 3), $25{,}048{,}311$ CHEB stars (label 4), and $10{,}651{,}805$ EAGB stars (label 7) are assigned to the corresponding best-fit template classes. All three classes correspond to relatively late and bright stages of stellar evolution and remain detectable by HST at the distance of M31. RGB stars are the most numerous and constitute the dominant tracer for the extinction measurements; CHEB stars are next in number, and being brighter than typical RGB stars, they trace larger distances or higher-extinction regions in luminosity. EAGB stars are the brightest of the three; though fewest in number, they can best trace sightlines through the highest extinction.

\begin{figure*}[htbp]
\centering
\includegraphics[width=\textwidth]{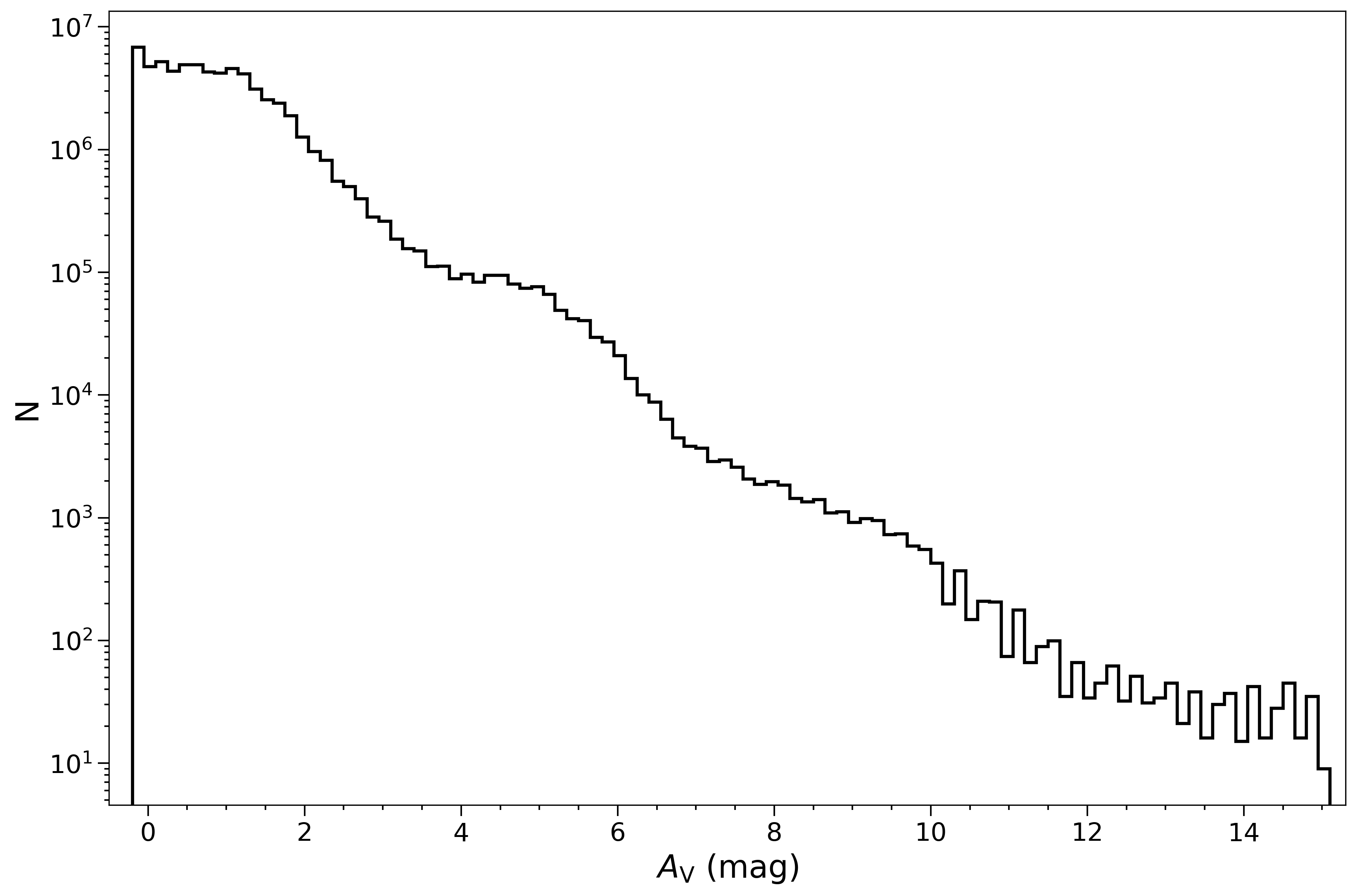}
\caption{Distribution of $A_V$ for the high-quality subsample ($Q\leq1$, 64,840,973 stars in total; bin width $0.15\,\mathrm{mag}$). The vertical axis shows the stellar count on a logarithmic scale.}

\label{distrav}
\end{figure*}

Figure~\ref{distrav} presents the $A_V$ distribution of the high-quality subsample. The distribution is concentrated mainly below $1.5\,\mathrm{mag}$ and extends toward higher extinction values, forming a pronounced right tail. \citet{Dalcanton2015} similarly found, in their CMD modeling of M31, that the dust column-density distribution within each region can be well described by a log-normal function. Approximately 97\% of the stars have $A_V<3.0\,\mathrm{mag}$, with only about 1.5\% exhibiting $A_V>4.0\,\mathrm{mag}$. The low fraction of highly extinguished stars may partly reflect observational selection effects, as stars along the most heavily extinguished sightlines are difficult to detect with HST due to severe extinction. Consequently, the observed stellar sample may underrepresent highly extinguished sightlines. This is consistent with the findings of \citet{Wang2025}, who reported extinction values of $3$--$4\,\mathrm{mag}$ in the densest spiral-arm regions of M31, with the same limitation imposed by source detectability in highly extinguished regions. By contrast, the extinction map of \citet{Draine2014}, based on dust emission, reaches $A_V\approx7\,\mathrm{mag}$ in some dense regions, reflecting the advantage that emission-based methods are not limited by source detectability, although emission photometry alone cannot independently verify the reliability of these high extinction values.

\subsection{The M31 Extinction Map}

\begin{figure*}[htbp]
\centering
\includegraphics[width=\textwidth]{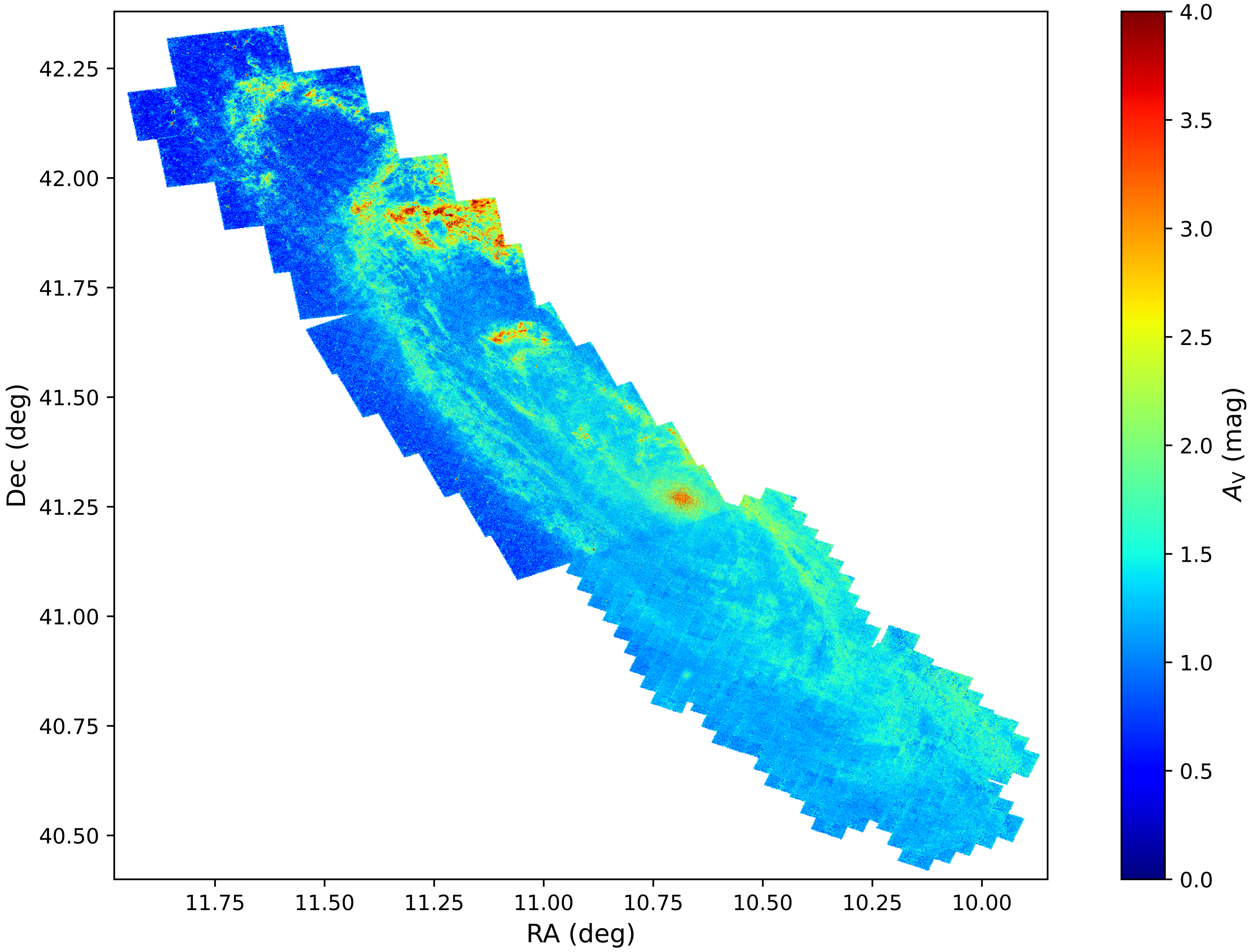}
\caption{Stellar-traced extinction map of M31 constructed in this work. The map has been publicly released through the NADC platform (\url{https://nadc.china-vo.org/res/r101847/}).}

\label{mapq}
\end{figure*}

Based on the high-quality subsample described above, the extinction map is constructed using the HEALPix adaptive-resolution scheme. The final extinction map is shown in Figure~\ref{mapq} and contains three adaptive-resolution levels: $6.44''$, $3.22''$, and $1.61''$. The highest resolution of $1.61''$ corresponds to a physical scale of approximately $6\,\mathrm{pc}$ at the distance of M31, which is about a factor of $4$ improvement over the $25\,\mathrm{pc}$ resolution of \citet{Dalcanton2015} and a factor of about $30$ improvement over the $50''$ ($\sim 189\,\mathrm{pc}$) resolution of \citet{Wang2025}. Such a high spatial resolution allows us to begin to resolve, in the extinction map of M31, structures on the scale of individual giant molecular clouds ($\sim 10$--$50\,\mathrm{pc}$), thereby providing a new observational basis for studying the small-scale density structure of the ISM.

The extinction map clearly reveals the prominent spiral-arm structure of M31 and is in close spatial agreement with known star-forming regions and HII regions. Quantitatively, within the northern region covered by PHAT, the extinction in the densest portions of the spiral arms reaches $A_V\approx2.0$--$4.0\,\mathrm{mag}$, while inter-arm regions and the outer disk drop below $0.5$--$1.0\,\mathrm{mag}$, yielding an extinction contrast of factors of 3--8. \citet{Wang2025} identify multiple spiral-arm groups located at radii of approximately 2.8, 5.6, 11.2, and $15.6\,\mathrm{kpc}$ in their extinction map; the present map shows a consistent spatial distribution of spiral arms in the corresponding PHAT region. Although the PHAST survey provides four optical and near-ultraviolet bands, most stars used for extinction measurements have only two valid photometric bands. Its large-scale spiral-arm structures remain clearly visible and appear to form a nearly continuous spiral pattern with the northern arms traced by PHAT, but its absolute extinction values and small-scale structures should be interpreted with caution. Together, the two surveys cover approximately two-thirds of the star-forming disk of M31. The median extinction across the entire map is $A_V=1.24\,\mathrm{mag}$, numerically close to the mean extinction of $A_V=1.17\,\mathrm{mag}$ reported by \citet{Wang2025}.

The extinction map developed in this work reveals a series of fine-scale structures that have been difficult to resolve in previous extinction maps of M31. First, filamentary network structures are observed between the two main spiral arms in the northern region. The typical widths of these filaments are several arcseconds (corresponding to $\sim 20$--$40\,\mathrm{pc}$), comparable to the scales of filamentary structures observed in nearby molecular clouds within the Milky Way. Numerical simulations and observational studies of ISM turbulence \citep[e.g.,][]{MacLow2004} indicate that supersonic turbulence naturally produces filamentary and sheet-like density structures during the energy cascade, and the filamentary networks observed here are consistent with this theoretical expectation. Second, multiple low-extinction cavities with typical sizes of tens of arcseconds are visible within the spiral-arm regions. The boundaries of these cavities frequently coincide with the locations of known HII regions or supernova remnants, suggesting that they may correspond to interstellar bubbles created by stellar winds and supernova explosions clearing the surrounding dust through mechanical-energy feedback. Whether these cavities are causally linked to specific feedback sources, however, still requires further dynamical evidence. Third, spur-like structures extending radially outward from the northern main spiral arms can be identified. These features are morphologically similar to the inter-arm material transport channels predicted by the spiral density-wave theory \citep{Lin1964}. Fourth, wavelike or feather-like periodic morphologies are observed between the main spiral arms in the central region. Theoretical studies indicate that such structures may be related to hydrodynamic instabilities within the spiral arms \citep{Wada2004}.

Once these fine-scale structures are quantitatively characterized, they will offer significant diagnostic value for dust physics. The topological properties of the filamentary network (such as the number of filaments per unit area and the power spectrum of filament length) may constrain the injection scale and energy spectral index of ISM turbulence, while the size distribution function of the cavities can reflect the efficiency of stellar feedback energy distribution across spatial scales. The extinction map produced in this work has been publicly released through the National Astronomical Data Center (NADC) platform (\url{https://nadc.china-vo.org/res/r101847/}), providing baseline data for systematic quantitative analyses of these structures. The map is also expected to provide high-resolution observational constraints for extinction correction and dust model improvement in the M31 region for future surveys such as the Chinese Space Station Survey Telescope (CSST; \citealt{CSST2026}).

\subsection{Comparison with Previous Studies}

\begin{figure*}[htbp]
\centering
\includegraphics[width=0.49\textwidth]{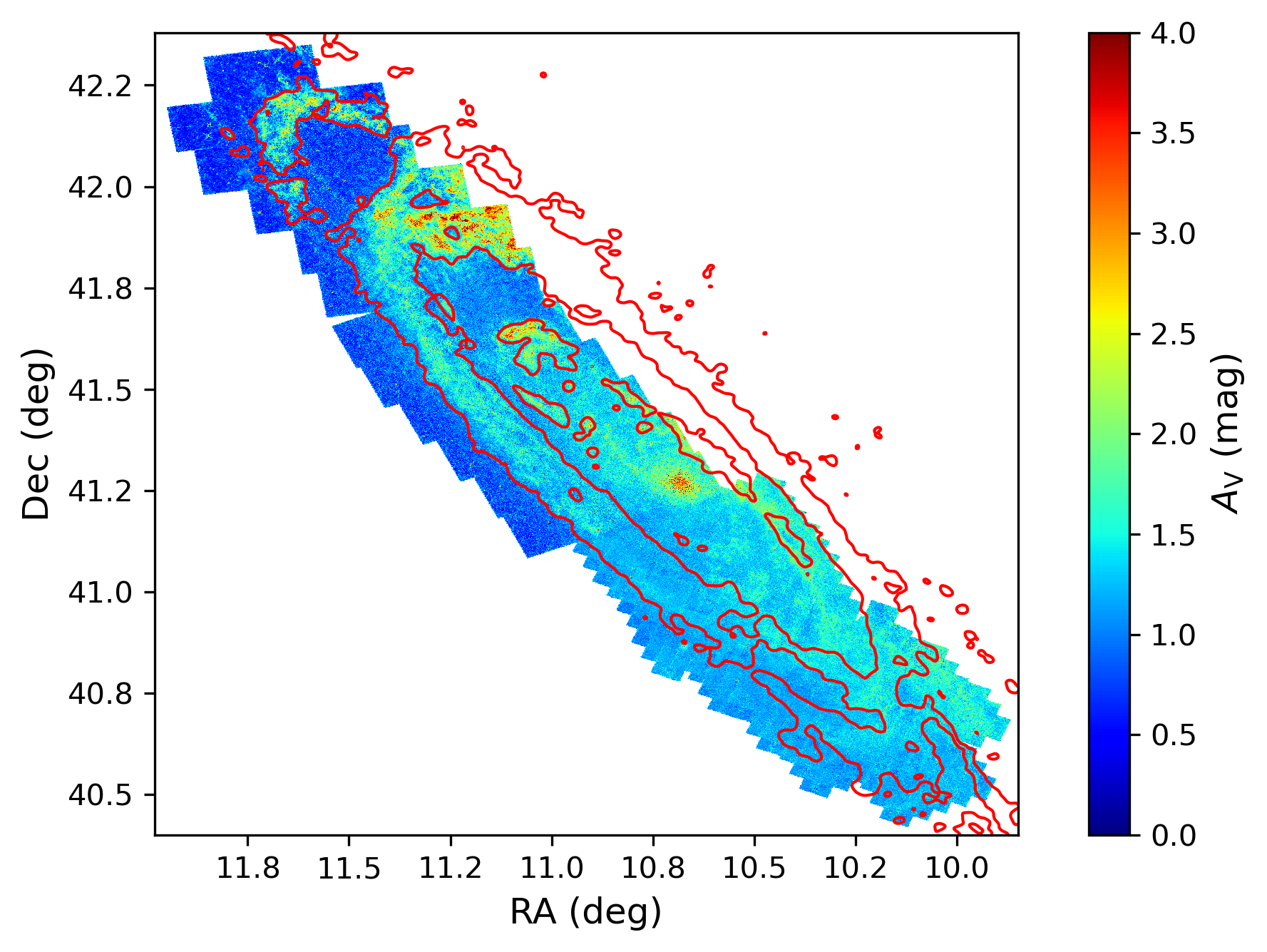}
\hfill
\includegraphics[width=0.49\textwidth]{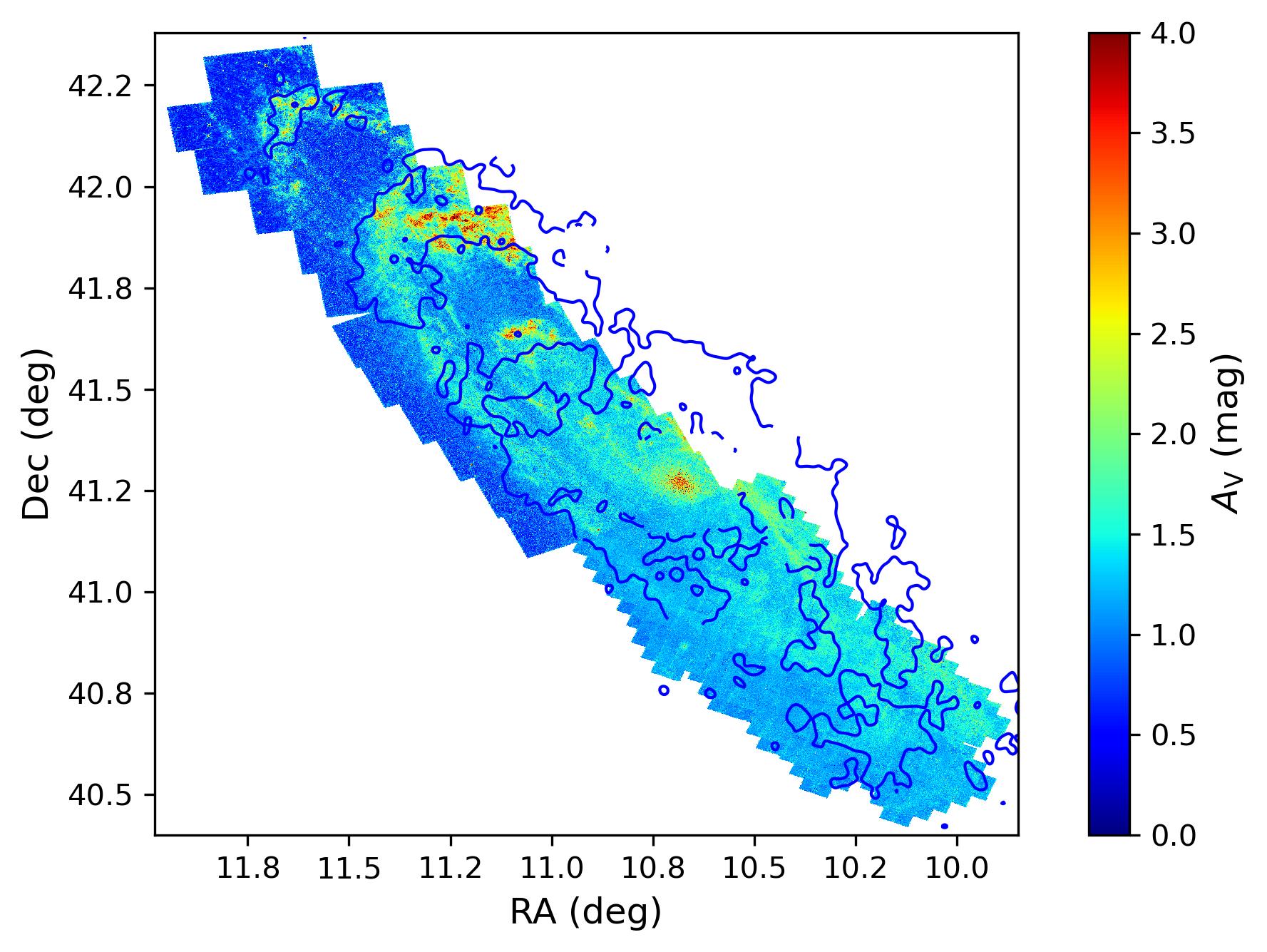}
\caption{Comparison of the large-scale morphology between the extinction map of this work (color background) and those of \citet{Draine2014} (left panel) and \citet{Wang2025} (right panel). The two panels use the same sky-coordinate range and background color scale. The red and blue contours indicate the $A_V=1.8\,\mathrm{mag}$ contour in the \citet{Draine2014} map and the $A_V=1.4\,\mathrm{mag}$ contour in the \citet{Wang2025} map, respectively. These contours are used to compare the morphology of the main spiral-arm structures.}

\label{mapwd}
\end{figure*}

Figure~\ref{mapwd} compares the extinction map of this work with those of \citet{Draine2014} and \citet{Wang2025}. The \citet{Draine2014} extinction map is based on Spitzer and Herschel infrared emission data and converts the dust mass surface density $\Sigma_{\rm Md}$ to $A_V$ through a dust radiative transfer model. The \citet{Wang2025} extinction map uses ground-based multi-band data from UKIRT/WFCAM, Pan-STARRS1, and Gaia DR3, with M31 member stars as tracers, and adopts a stellar color-locus matching method to derive the extinction independently. Although the three maps are based on entirely independent technical approaches, they agree well on large scales in showing the main spiral-arm structure of M31.

\begin{figure*}[htbp]
\centering
\includegraphics[width=0.9\textwidth]{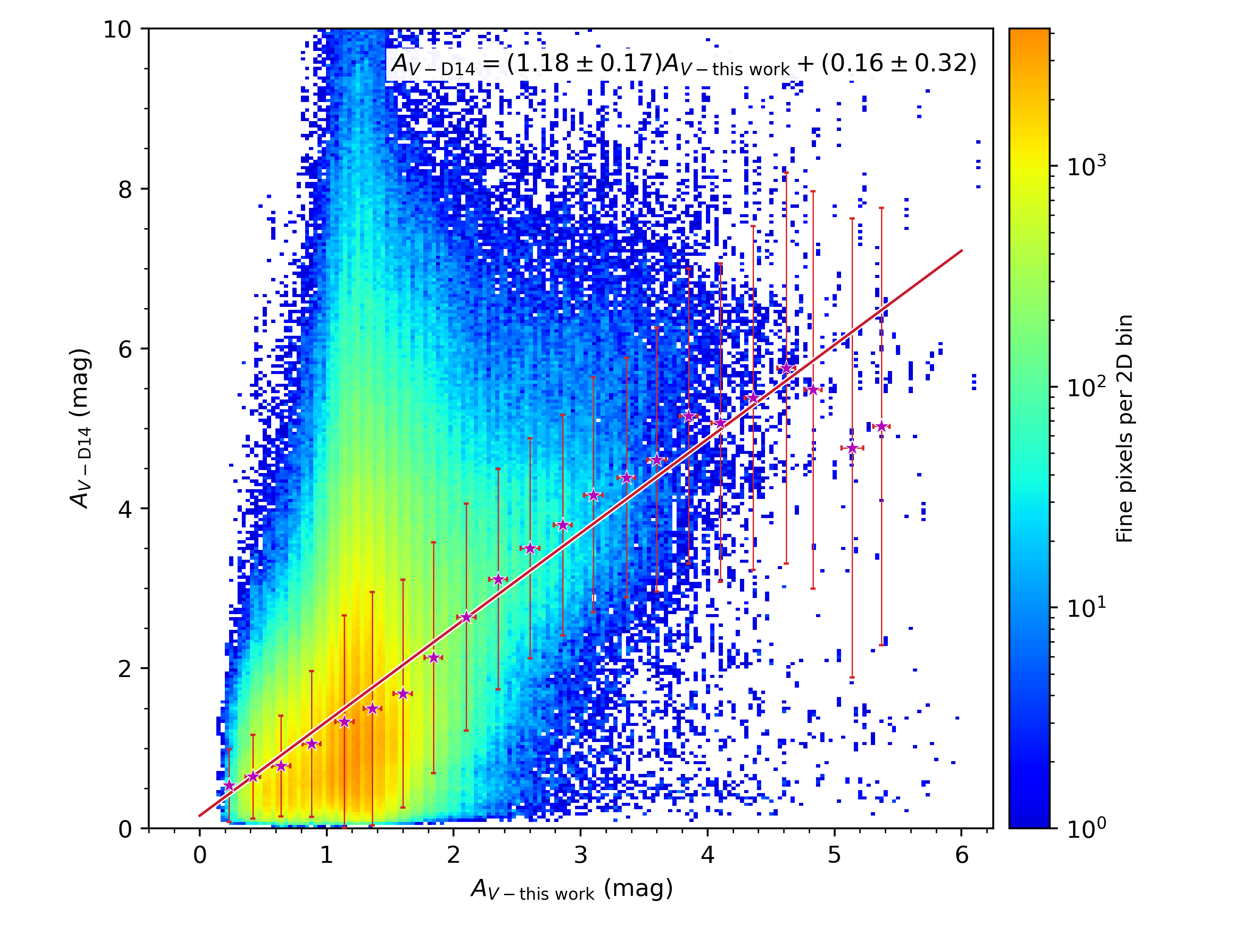}
\caption{Position-by-position comparison between the extinction map derived in this work and that of \citet{Draine2014}, excluding the central $3'$ region of M31. The colored density distribution shows the number of matched positions in each two-dimensional bin on a logarithmic scale. Magenta stars indicate the median values after binning the matched pixels by $A_V$ from this work with a bin width of $0.25\,\mathrm{mag}$, retaining only bins containing more than 50 matched pixels; the vertical error bars represent the standard deviation of the \citet{Draine2014} extinction values within each bin. The red line shows the best-fitting linear relation to the binned medians obtained with MCMC,
$A_V^{\rm(D14)}=(1.18\pm0.17)A_V^{\rm(this\ work)}+(0.16\pm0.32)$.}

\label{mapcompare}
\end{figure*}

Figure~\ref{mapcompare} presents a direct position-by-position quantitative comparison between the extinction values derived in this work and those of \citet{Draine2014}. After excluding the central $3'$ region of M31, we bin the matched positions by $A_V$ from this work using a bin width of $0.25\,\mathrm{mag}$ and retain only bins containing more than 50 matched pixels. A linear fit to the binned medians yields $A_V^{\rm(D14)}=(1.18\pm0.17)A_V^{\rm(this\ work)}+(0.16\pm0.32)\,\mathrm{mag}$. Tests using bin widths of $0.125$--$0.50\,\mathrm{mag}$ and a shifted $0.25\,\mathrm{mag}$ binning scheme give best-fitting slopes of approximately $1.16$--$1.20$. The binned medians show an approximately linear overall relation, although substantial scatter is present in the position-by-position comparison. Therefore, the fitted relation should not be interpreted as a fixed scaling factor between the two maps. \citet{Dalcanton2015} used a near-infrared CMD modeling method based on the PHAT survey to construct a $25\,\mathrm{pc}$-resolution extinction map and reported a systematic offset of approximately a factor of $2.5$ between their values and those of \citet{Draine2014}, with the \citet{Draine2014} values being systematically higher. They attributed this offset to an underestimate of the dust grain emissivity of M31 in the dust model of \citet{Draine2007}. \citet{Wang2025}, using ground-based multi-band data, reported a scaling factor of approximately $1.28$ relative to \citet{Draine2014}. \citet{Dalcanton2015} uses a CMD-modeling approach based on a statistical model assuming a log-normal dust column-density distribution; this work uses a direct matching method based on theoretical stellar templates; and \citet{Wang2025} uses a stellar color-locus intersection method. The CMD-modeling approach is sensitive to assumptions about the functional form of the dust column-density distribution and to the foreground/background star fraction parameter $f_{\rm red}$, whereas the stellar template-matching method directly fits individual stars and reduces the introduction of such statistical assumptions. \citet{Dalcanton2015} also note that the magnitude of the offset between \citet{Draine2014} and their extinction values is correlated with the hardness of the local interstellar radiation field: in regions with harder radiation fields, the overestimate by the \citet{Draine2014} model is smaller, suggesting that the emissivity bias may be physically associated with the heating efficiency of the radiation field. Because the \citet{Dalcanton2015} extinction map has not yet been publicly released, a direct pixel-by-pixel comparison with our values is not currently possible.

\begin{figure*}[htbp]
\centering
\includegraphics[width=0.95\textwidth]{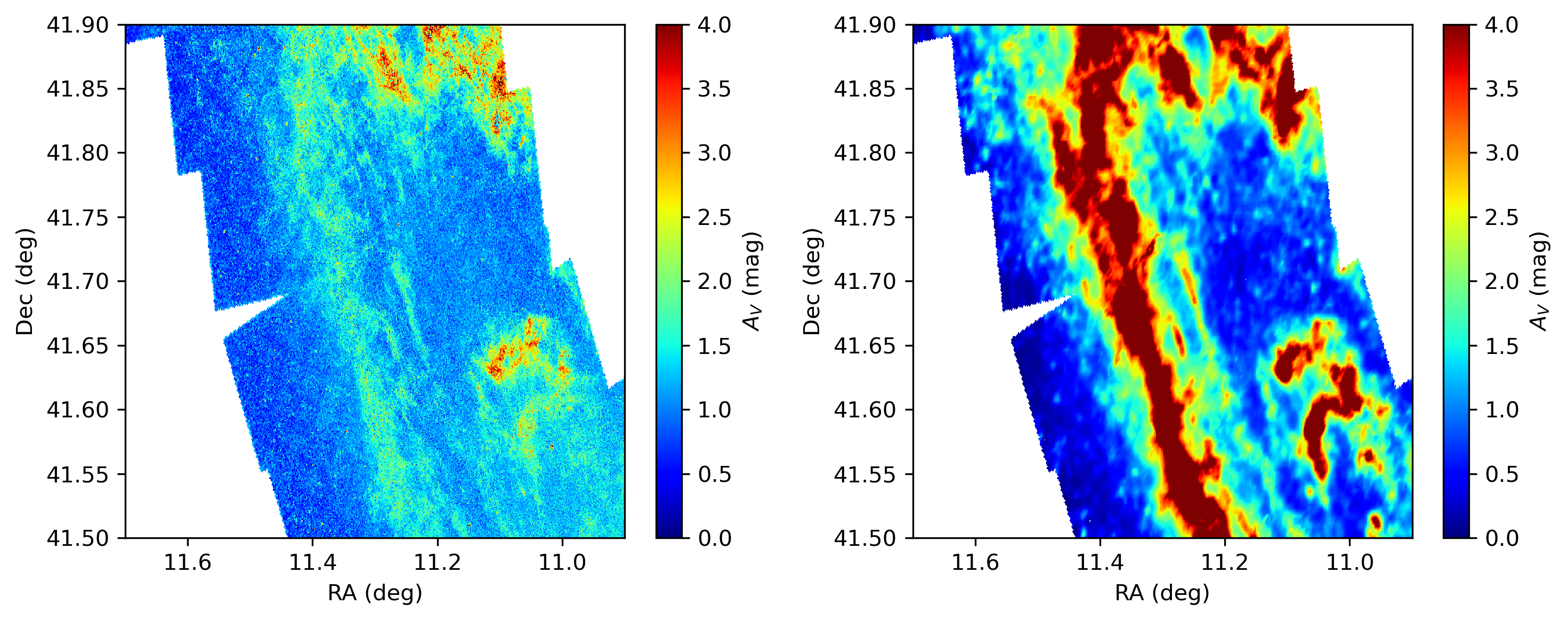}
\\
\vspace{4ex}
\includegraphics[width=0.95\textwidth]{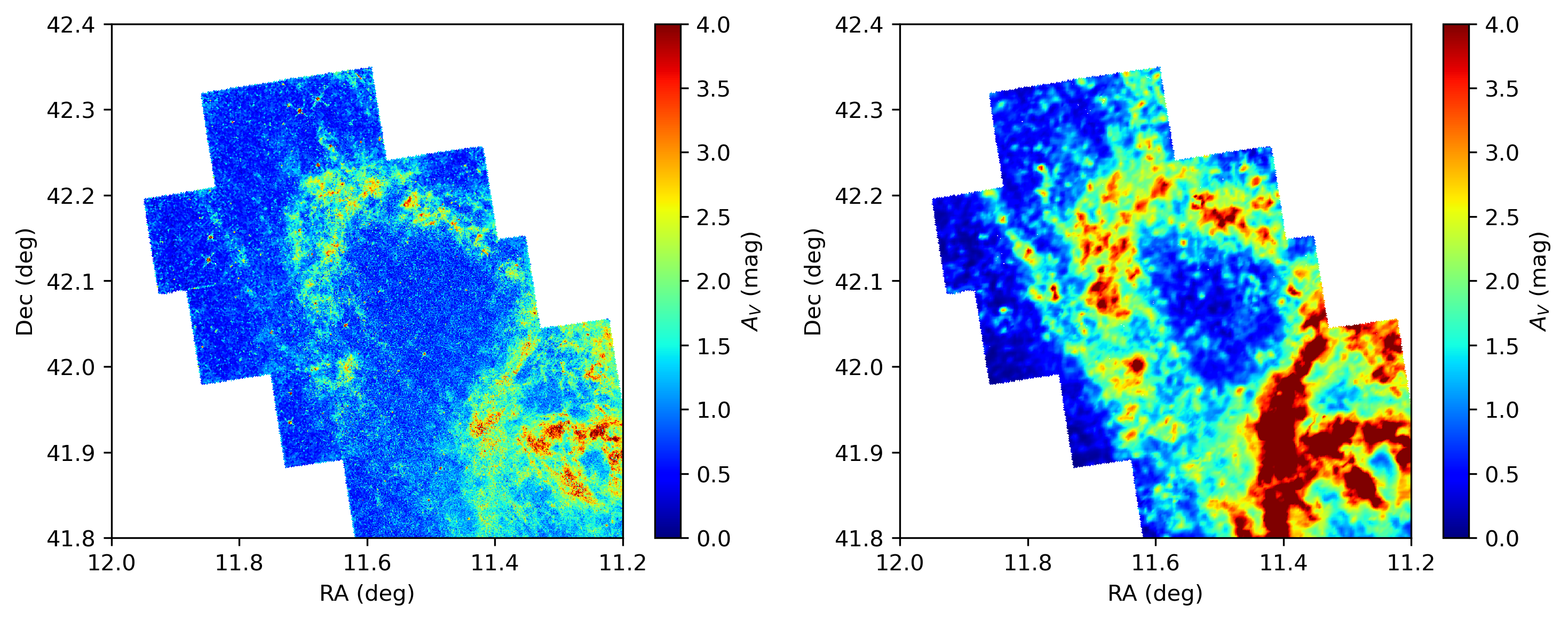}
\caption{Local morphological comparison between the extinction map of this work and the dust-emission-based extinction map of \citet{Draine2014} in two selected regions. The upper panel corresponds to Region 1 (RA $\in [10.9^\circ, 11.7^\circ]$, Dec $\in [41.5^\circ, 41.9^\circ]$), located in the northern main spiral arm of M31; the lower panel corresponds to Region 2 (RA $\in [11.2^\circ, 12.0^\circ]$, Dec $\in [41.8^\circ, 42.4^\circ]$), covering another segment of the northern main spiral arm and its inter-arm region. In each subpanel, the left side shows the extinction map of this work (adaptive resolution, up to $1.61''$), while the right side shows the \citet{Draine2014} map (effective angular resolution $\sim24.9''$). All panels use the same color scale.}

\label{mapj}
\end{figure*}

Figure~\ref{mapj} further compares the detailed morphology of the extinction maps of this work and \citet{Draine2014} in two selected regions. On large scales, the two maps show similar major morphological features, including the spiral-arm structures. However, substantial local differences remain: in some regions where the extinction in this work is approximately $A_V\simeq2\,\mathrm{mag}$, the corresponding values in the \citet{Draine2014} map can reach $A_V>4\,\mathrm{mag}$, indicating that the numerical differences between the two maps cannot be described by a single scaling factor. These differences may be jointly affected by spatial resolution, assumptions in the dust-emission modeling, and selection effects in the detectable stellar population. In addition, owing to the high spatial resolution of $1.61''$, our extinction map reveals fragmented filamentary networks and small-scale cavities in the inter-arm and outer-arm regions that are unresolved in the \citet{Draine2014} map. The spatial scales of these fine structures ($\sim10$--$50\,\mathrm{pc}$) are smaller than the effective resolution of the \citet{Draine2014} extinction map used in this comparison ($\sim24.9''$, corresponding to approximately $94\,\mathrm{pc}$ at the distance of M31), demonstrating the value of high-resolution stellar extinction maps for studying small-scale ISM structures.

In summary, the extinction map constructed in this work provides an independent observational reference for constraining dust grain models in M31. A central limitation of dust-emission-based extinction maps, such as that of \citet{Draine2014}, is that the absolute extinction calibration depends on assumptions about the dust-emission properties. Stellar extinction maps trace dust extinction through the attenuation of background starlight and do not depend on the emission properties of the dust grains, thereby providing an independent reference for comparison with emission-based models. However, stellar extinction maps may also be affected by source detectability and sample-selection effects, so the numerical differences between the two types of extinction maps cannot be uniquely attributed to dust emissivity. Future systematic comparisons of these differences with environmental parameters such as the local radiation-field hardness, metallicity, and dust temperature may help establish empirical constraints on the environmental dependence of dust-emission properties in M31. Such constraints may further help us understand how the size distribution and chemical composition of dust grains vary with galactic environment and provide observational guidance for the development of dust models.

\subsection{Discussion of Special Structures}\label{sec:special}

Several morphological features and systematic effects in the extinction map also deserve further discussion.

\begin{figure*}[htbp]
\centering
\includegraphics[width=\textwidth]{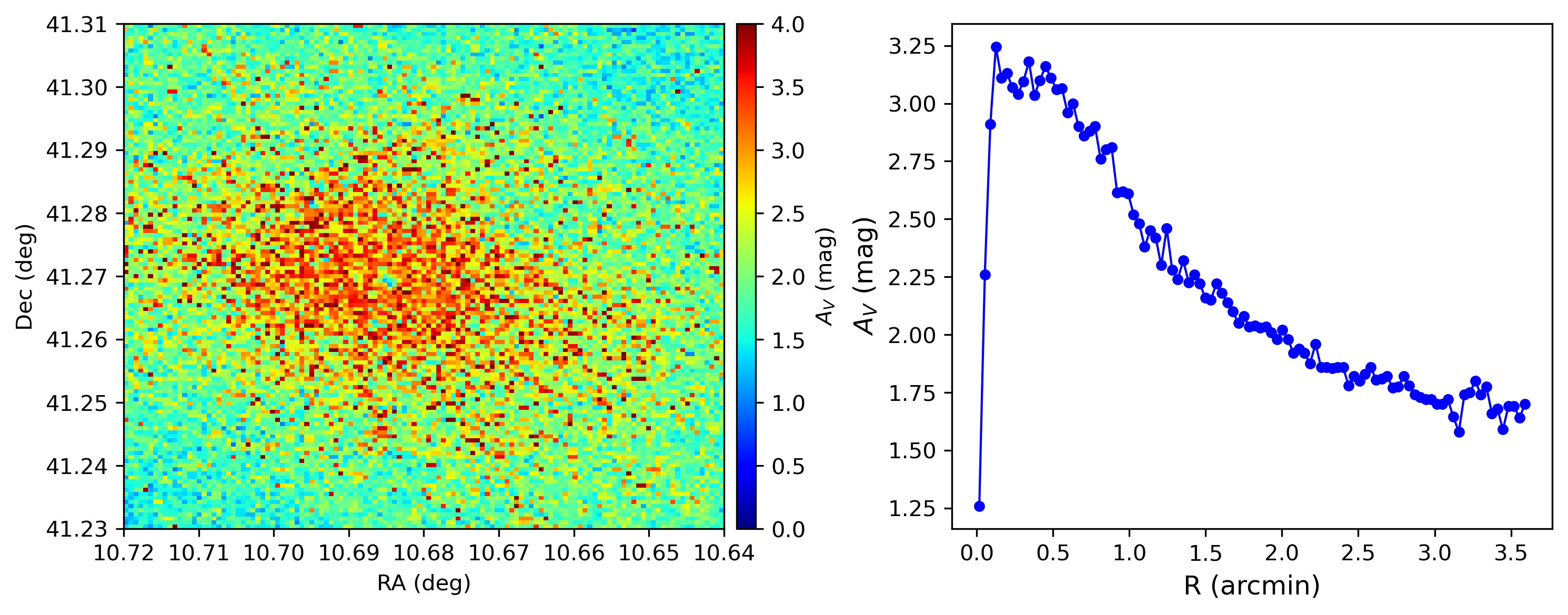}
\caption{Detailed view of the extinction structure in the central region of the M31 bulge. Left: spatial distribution of $A_V$ in this region. Right: variation of the median $A_V$ with radial offset $R$ from the adopted center. The radial offset $R$ is shown in arcminutes, and each point represents the median $A_V$ within the corresponding radial bin.}

\label{zxq}
\end{figure*}

\begin{figure*}[htbp]
\centering
\includegraphics[width=\textwidth]{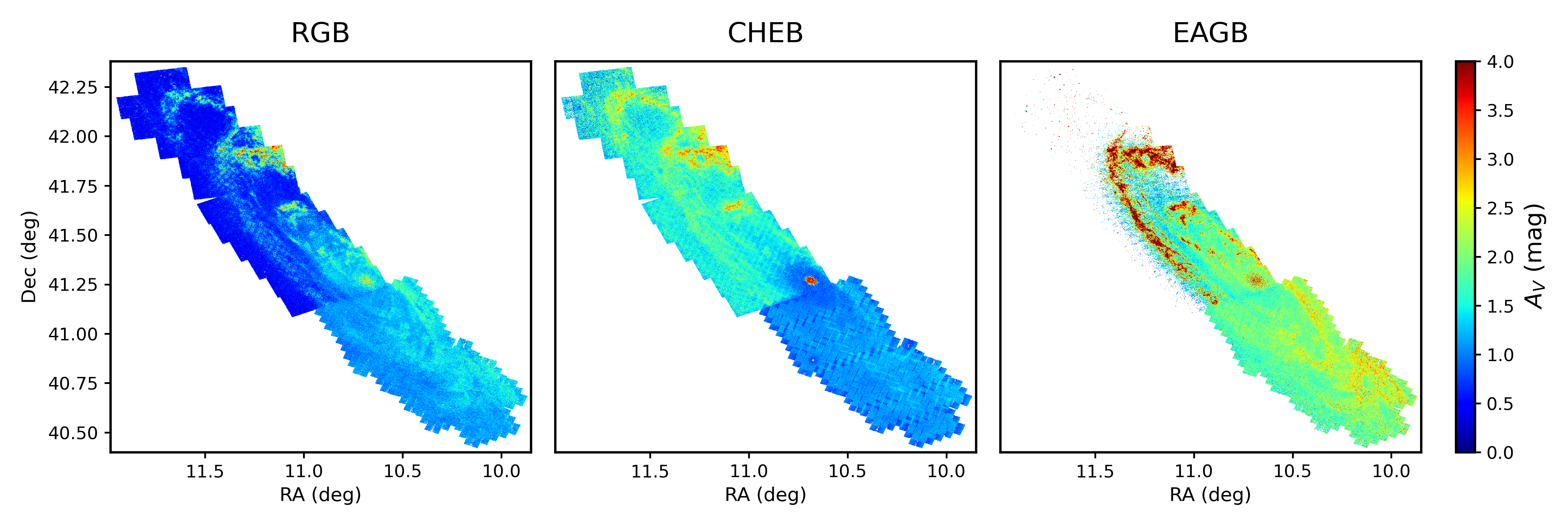}
\caption{Extinction maps of M31 constructed using different stellar types as extinction tracers. The construction methodology (adaptive resolution, pixel-by-pixel median combination) is identical to that of Figure~\ref{mapq}. Left: RGB stars (PARSEC label 3, approximately 45\% of the $Q\leq1$ sample). Middle: CHEB stars (core helium burning, label 4, approximately 39\%). Right: EAGB stars (early asymptotic giant branch, label 7, approximately 16\%).}

\label{mtype}
\end{figure*}

Central spherical structure. In the central region of M31, the extinction map detects a prominent spherical extinction enhancement structure spanning approximately $3$ arcminutes (corresponding to about $680\,\mathrm{pc}$). Its morphology differs significantly from the features in the overlap region of the high-resolution $180'' \times 180''$ extinction map obtained by \citet{Dong2016} from HST ten-band SED fitting. The left panel of Figure~\ref{zxq} provides a detailed view of this region, while the right panel shows the radial variation of $A_V$ from the center of the structure. The innermost part of the structure is characterized by low extinction ($A_V \approx 1.25\,\mathrm{mag}$), consistent with the established understanding that the central bulge region of M31 contains relatively little dust, given that the bulge is dominated by old stellar populations and the local ISM density is relatively low. However, on scales of only about $0.1$ arcminutes from the center, the extinction rises sharply to about $3.25\,\mathrm{mag}$, forming a steep extinction gradient. Beyond a radius of about $0.6$ arcminutes, the extinction gradually decreases to the background level. The extinction traced by different stellar types (RGB, CHEB, and EAGB) consistently exhibits this ring-like extinction enhancement (Figure~\ref{mtype}), indicating that the feature does not depend on the selection of any specific stellar evolutionary phase.

Two possible explanations for the physical origin of this structure can be considered. The first explanation is that the feature is a photometric artifact: the stellar surface density in the central region is extremely high, particularly in the WFC3/IR field of view, where the source density reaches $\sim 10^4$--$10^5\,\mathrm{stars}\,\mathrm{arcmin}^{-2}$, leading to severe stellar overlap and crowding effects in the DOLPHOT photometry. In crowded star fields, the photometric flux from two or more neighboring stars can be incorrectly assigned to a single source, producing a spurious ``brightening'' that feeds through the $\chi^2$ fit into a systematic overestimate of the extinction. This interpretation is consistent with the choice of \citet{Dalcanton2015} to skip Brick 1 in their analysis of the M31 bulge region (because the high surface brightness causes severe crowding that compromises the CMD-based extinction modeling) and with the treatment strategy of \citet{Dong2016} for this region in their high-resolution extinction map. The second explanation is that the feature represents a real astrophysical structure: the Central Molecular Zone (CMZ) of the Milky Way shows a similar ring-like distribution of dust and gas \citep{Molinari2011}, and the bulge region of M31 may also harbor a dust ring or gas disk on a similar scale. In addition, the bulge of M31 is known to contain a $\sim 1\,\mathrm{kpc}$ bar structure \citep{Beaton2007}, and gas dynamical processes within the bar (such as gas transport and accumulation along the bar) can naturally produce a ring-like extinction enhancement. Nevertheless, the structure shows a steep extinction gradient across a very small spatial scale ($\sim 0.1\,\mathrm{arcmin} \approx 23\,\mathrm{pc}$), which, if real, would imply an unusually steep dust column-density gradient requiring extreme gas-compression mechanisms. Considering the available evidence, although the possibility of a real dust structure cannot be entirely ruled out, the photometric-artifact interpretation associated with crowded star fields is the simpler explanation. Distinguishing between these possibilities will require future observations with higher angular resolution or improved crowded-field photometry techniques. In the publicly released extinction map, users are advised to treat the extinction values within this central region (within a radius of approximately $3$ arcminutes) with appropriate caution.

Differences in extinction traced by different stellar types. Identifiable differences are present in the extinction structures traced by stars in different evolutionary phases, providing valuable information for understanding the systematic uncertainties in the extinction measurements. The extinction map traced by RGB stars exhibits a relatively low average $A_V$, with clearly visible spiral arms in both the northern and southern regions and a relatively large central spherical region characterized by relatively low extinction. The extinction traced by CHEB stars shows asymmetry between the northern and southern regions: the overall extinction level is higher in the north than in the south, and the spiral-arm structure remains identifiable in the north but is less distinct in the south. In addition, the outer part of the central spherical region shows very low extinction, while the innermost part exhibits high extinction. The extinction traced by EAGB stars has the highest average $A_V$ among the three types, with the most prominent spiral-arm structures and the most pronounced extinction contrast in the northern region. The broad similarity in the large-scale morphology traced by the three stellar types supports the overall extinction morphology; however, their quantitative differences should be interpreted with caution. Stars of different luminosities probe dust layers to different depths. Brighter EAGB stars can remain detectable through larger dust columns and therefore have a higher probability of tracing background stellar populations behind high extinction, resulting in a higher average $A_V$. Fainter RGB stars have lower completeness along heavily extinguished sightlines and tend to sample lower-extinction sightlines. This observational selection effect is an intrinsic feature of stellar extinction maps, which reflect the extinction distribution of the detectable stellar population rather than a complete sampling of all sightlines. In addition, systematic differences between stellar models and observed stellar SEDs may introduce type-dependent extinction biases, so the quantitative differences among stellar types cannot be attributed solely to observational selection effects. The potential influence of this selection effect and possible model-related systematics on studies of star formation history and on estimates of the ISM mass should be carefully considered in future applications of the extinction map.

We further examined the observed-minus-model magnitude residuals as a function of fitted temperature and stellar type. Among the bands with sufficient statistics, F110W shows the strongest dependence on stellar population: the mean residuals are $+0.0121$ and $-0.0162\,\mathrm{mag}$ for stars with fitted $T_{\rm eff}<5000\,\mathrm{K}$ and $T_{\rm eff}\geq5000\,\mathrm{K}$, respectively, while the temperature-dependent differences are smaller in the other well-sampled bands. F110W also shows the clearest dependence on the best-fitting stellar type, whereas the differences are generally smaller in the other well-sampled bands. These internal fitting residuals suggest possible stellar-population-dependent systematic effects, although their origin and quantitative impact on $A_V$ cannot be determined from the residuals alone.

Systematic differences between the PHAST and PHAT regions. The spiral-arm features in the extinction map of the PHAST region are noticeably less distinct than in the PHAT region, and the boundary between the two regions can be discerned in the map. The photometric information provided by the two surveys differs substantially. The PHAST dataset nominally includes four bands (F275W, F336W, F475W, and F814W), but in practice 99.8\% of the sources have measurements in only two of these bands. This implies that the extinction measurement in the PHAST region effectively reduces to a two-band color match, in which extinction and temperature are strongly degenerate and cannot be uniquely determined. The PHAT dataset, by contrast, includes six bands from F275W to F160W, with 36.2\% of the sources having independent measurements in four or more bands; the addition of the infrared bands (F110W and F160W) provides effective dispersion between the extinction and temperature sensitivities, helping to better constrain both. This finding offers a clear empirical reference for the design of observational strategies in similar future surveys: even if the sensitivity in the near-infrared bands is limited, the inclusion of one or two near-infrared filters can substantially improve the constraints on extinction measurements, rather than merely improving their statistical precision. With respect to the current data, the PHAST extinction map should be used primarily to examine large-scale relative extinction morphology, while its absolute extinction values and small-scale structures require caution in quantitative applications. The base resolution of $nside=2^{15}$ ($6.44''$) may be used to reduce random fluctuations through spatial averaging, although this cannot eliminate systematic uncertainties associated with the two-band degeneracy. In addition, grid-like structures resembling the PHAT and PHAST survey tiling are visible in the extinction map (Figure~\ref{mapq}), particularly in low-extinction regions. These structures may be related to differences in survey coverage, residual field-to-field photometric calibration, and sample-selection effects arising from variations in exposure depth and source detectability. Therefore, these patterns should not be directly interpreted as real small-scale dust structures.

\section{Summary}\label{sec:sum}

In this study, a high-resolution (up to $1.61''$, $\sim 6\,\mathrm{pc}$) and large-coverage (approximately two-thirds of the star-forming disk) stellar-traced extinction map of M31 is constructed based on multi-band HST photometric data from the PHAT and PHAST surveys, and has been publicly released through the NADC platform. The median extinction across the map is $A_V=1.24\,\mathrm{mag}$, numerically close to the mean extinction of $A_V=1.17\,\mathrm{mag}$ reported by \citet{Wang2025} based on ground-based data.

In terms of ISM physics, the high-resolution extinction map identifies fine-scale features such as filamentary networks, low-extinction cavities, spur-like structures, and feather-like periodic morphologies, providing a quantitative observational basis for the study of physical processes such as the energy cascade of ISM turbulence, the distribution of stellar feedback energy, and hydrodynamic instabilities within spiral arms. In addition, in terms of dust grain physics, the comparison with \citet{Draine2014} reveals substantial differences between the two extinction maps, which may reflect both uncertainties in the assumed dust grain emissivity and selection effects in the detectable stellar population. Stellar extinction maps provide an independent observational reference based on the attenuation of background starlight for comparison with dust-emission models. This work also reveals two important systematic limitations of the stellar-extinction-map method. First, stars of different luminosities probe the dust layer to different depths (EAGB $>$ CHEB $>$ RGB), so the extinction map essentially reflects the extinction distribution of the detectable stellar population rather than a complete accounting of all sightlines. In addition, systematic differences between stellar models and observed stellar SEDs may introduce stellar-type-dependent extinction biases. Second, multi-band coverage is essential for reducing the degeneracy between extinction and temperature: although PHAST nominally includes four photometric bands, most stars in this region have only two valid bands, leading to a strong temperature--extinction degeneracy. The extinction estimates in PHAST are therefore less reliable than those in PHAT, and their absolute values and small-scale structures should be interpreted with caution. These two points carry important implications for future applications of stellar extinction maps and for the design of survey strategies. The ring-like extinction enhancement structure in the central region of the M31 bulge is more likely to be caused by photometric artifacts in crowded star fields, and users are advised to exercise caution within this region (within a radius of approximately $3'$).

The extinction map provides a key high-resolution baseline for extinction correction in future surveys such as CSST in the M31 region \citep{CSST2026}, offers quantitative constraints for studies of ISM turbulence, stellar feedback, and dust evolution, and supports subsequent scientific studies of M31.

\section*{Acknowledgments}
This work is partially supported by the National Natural Science Foundation of China No. 12203016, 12322304, 12303030 and 12573026, the Natural Science Foundation of Hebei Province No. A2022205018, the Shandong Provincial Natural Science Foundation No. ZR2024QA236, the Science Research Project of the Hebei Provincial Department of Education No. BJ2026095 and No. CYZD2026002, the National Natural Science Foundation of Yunnan Province No. 202301AV070002, the Xingdian Talent Support Programme of Yunnan Province, and the Innovation and Entrepreneurship Training Program for College Students of Hebei Normal University No. 202510094027. W.Y.C. acknowledges support from the China Manned Space (CMS) Program research project. M.X.S. acknowledges support from the Postdoctoral Research Station in Physics at Hebei Normal University. We acknowledge the support of the China Manned Space Program research project through the Chinese Space Station Survey Telescope No. CMS-CSST-2025-A11.

\bibliographystyle{aasjournal}
\bibliography{spp}

\end{CJK*}
\end{document}